\documentclass[showpacs,twocolumn,amsmath,amssymb]{revtex4}

\usepackage{graphicx}

\usepackage{dcolumn}% Align table columns on decimal point
\usepackage{bm}% bold math

\begin{document}

\title{Electrical manipulation of an electronic two-state system in Ge/Si quantum dots}
\author{C. E. Pryor}
 \email{craig-pryor@uiowa.edu}
\author{M. E. Flatt\'e}
\affiliation{
Optical Science and Technology Center and Department of Physics and Astronomy, University of Iowa, Iowa City, Iowa, 52242, USA}

\author{J. Levy}
\affiliation{
Department of Physics and Astronomy, University of Pittsburgh, Pittsburgh PA}

\date{\today}

\begin{abstract}
We calculate that the electron states of strained self-assembled Ge/Si quantum dots provide a convenient two-state system for electrical control. An electronic state localized at the apex of the quantum dot is nearly degenerate with a state localized at the base of the quantum dot. Small electric fields shift the electronic ground state from apex-localized to base-localized, which permits sensitive tuning of the electronic, optical and magnetic properties of the dot. As one example, we describe how spin-spin coupling between two Ge/Si dots can be controlled very sensitively by shifting the individual dot's electronic ground state between apex and base.
\end{abstract}

%73.      Electronic structure and electrical properties of surfaces,
%  73.63.Kv Quantum dots
% 03.67.Lx Quantum computation


\maketitle

{\it Introduction}.---Confinement of electrons in semiconductor quantum dots allows the energy, optical response, and spin response of the electron to be controlled through design of the structure and the controlled application of electric fields\cite{Bimberg1998,Hanson2007}. Material composition, strain, and structure all contribute to the confining potential of the dot, which can differ greatly for electrons and holes. Electric fields can confine single electrons to quantum dots in group-IV and III-V semiconductors, but only at very low temperature\cite{Ciorga2000,Hanson2007,Simmons2007}. Single electrons can be confined to single dots at room temperature, but only if the dominant confining potential is from structure and composition, not from electric fields. Thus one might expect only small changes in ground state properties to be achievable at room temperature with an applied electric field.

We report here that the ground state properties of electrons confined to Ge/Si quantum dots at room temperature can be sensitively tuned with electric fields due to the unusual near-degeneracy of two classes of electronic states. The lowest electronic states of Ge/Si quantum dots are confined predominately in the silicon surrounding the germanium\cite{Schmidt.prb.2000}. We find that the lowest energy states are at the apex (apex-localized-states, or ALS) and the next lowest energy states are at the base (base-localized states, or BLS). The energies of these two classes of states, ALS and BLS, are similar enough that a moderate electric field can change the ground state from an ALS to a BLS. An electron confined to a Ge/Si quantum dot thus forms a simple single-electron switch, and all ground-state properties of the dot should be very sensitive to the applied electric field.  
After presenting our calculations of the ALS and BLS states, and their electric-field response, we analyze the nature of the spin-spin coupling between neighboring quantum dots, and show that the exchange coupling between two ALS or two BLS states can be substantial, whereas the ALS-BLS coupling is minimal. 
For spin-based computation\cite{Loss1998,Kane1998}, and especially for quantum computation using Ge/Si quantum dots\cite{Levy2001}, this permits control of two-qubit gates much more sensitively than by changing the exchange constant by changing a state's binding energy and thus its wavefunction decay length.

{\it Method}.---We have calculated the electronic states of Ge/Si quantum dots using a strain-dependent four-band {\bf k}$\cdot$ {\bf p} theory for the valence states\cite{Pryor1997}, and multi-valley anisotropic effective mass theory for the conduction states.  The conduction states were calculated using an anisotropic effective mass theory with three components, one for each direction in which the $\Delta$-valleys occur. Due to strain, the band edge for a conduction band valley in the $\hat i$-direction undergoes a strain-induced shift given by
\begin{eqnarray}\label{HDstrain}
\Delta E_c = \Xi_d~ e + \Xi_u ~e_{ii}
\end{eqnarray}
where $e_{ii}$ is the diagonal element of the strain tensor, and $e = \sum_i ~ e_{ii} $ is the hydrostatic strain.

The strain due to the lattice mismatch between Si and Ge was calculated using continuum elasticity and cube shaped linear finite elements, with one element per unit cell of the crystal\cite{Pryor1997}.   The strain was then used as input to the calculation of electronic bound states in and around the island, using the envelope approximation on a real-space grid, with energies and wave functions computed using the Lanczos algorithm.

The Hamiltonian for the conduction states is given by 
\begin{eqnarray}\label{HD1}
H = \left(\begin{array}{ccc}H_x  & 0 & 0 \\0 & H_y  & 0 \\0 & 0 & H_z \end{array}\right)
\end{eqnarray}
where each of the $H_i$'s are defined by
\begin{eqnarray}\label{HD2}
H_{ x} \psi({\bf  r} ) &=&  
\frac{\hbar^2}{2\epsilon^2 m_{\|} }
\left[
 2 \psi({\bf  r} ) -\psi({\bf  r}  + \epsilon \hat  {\bf x} ) - \psi( {\bf  r} - \epsilon \hat {\bf  x} )
\right] \nonumber \\
&+&\frac{\hbar^2}{2\epsilon^2 m_{\perp} }
\left[
  2 \psi({\bf  r} ) -\psi({\bf  r} + \epsilon \hat  {\bf y} ) - \psi( {\bf  r} - \epsilon \hat  {\bf y} )
\right] \nonumber \\
&+&\frac{\hbar^2}{2\epsilon^2 m_{\perp} }
\left[
  2 \psi({\bf  r} ) -\psi({\bf  r} + \epsilon \hat  {\bf z} ) - \psi( {\bf  r} - \epsilon \hat  {\bf z} )
\right]\nonumber \\
&+& V({\bf  r }) \psi({\bf  r} )
 \end{eqnarray}
where ${\bf  r}$ are the locations of grid sites, $\epsilon$ is the grid spacing (taken to be equal to the lattice constant for Si), and $V({\bf r})$ includes the strain induced shift given by Eq. \ref{HDstrain}.
The longitudinal and transverse effective masses $m_{\|}$ and $m_{\perp}$ are given in Table \ref{materialParams}. Valleys related by spatial inversion (e.g. [100] and [$\overline{1}$00]) are treated as equivalent.
\begin{table}[h]
\begin{ruledtabular}
\begin{tabular}{ccc}
Parameter  &  Si  &  Ge \\
\hline
$E_g$       & 1.170 {\rm eV} & 0.740 {\rm eV}     \\% $\Delta_{so}$ (eV) & 0.0441 & 0.295\\
$E_v$       & 0  {\rm eV} & 0.370 {\rm eV}  \\

$\gamma_1$ & 4.26 & 13.35\\
$\gamma_2$ & 0.38 & 4.25\\
$\gamma_3$ & 1.56 & 5.69\\

$a_v$ & -2.1  {\rm eV}~\tablenotemark[1]  & 2 {\rm eV} \\
$b$    & -2.2  {\rm eV} & -2.9 {\rm eV} \\
$d$    & -5.1  {\rm eV} & -5.3 {\rm eV} \\

$a_{latt}$  & 0.5431  {\rm nm} & 0.5657  {\rm nm} \\
$C_{xxxx}$ & 1666  {\rm GPa} & 1290 {\rm GPa}\\
$C_{xxyy}$ & 640 {\rm GPa}& 480 {\rm GPa}\\
$C_{xyxy}$ & 800 {\rm GPa} & 670 {\rm GPa}\\

$\Xi_u$ & 10.5   {\rm eV}~\tablenotemark[1]  & 16.8  {\rm eV} ~\tablenotemark[1] \\
$\Xi_d$ & 1.1   {\rm eV}~\tablenotemark[1]  & -4.43   {\rm eV} ~\tablenotemark[1]\\

$m_{\|}$ & $0.1905 ~m_0$& -\\
$m_{\perp}$ & $0.9163 ~m_0$ & -\\

\end{tabular}
\end{ruledtabular}
\tablenotetext[1]{ Reference\ \onlinecite{Fischetti.jap.1996} }
\caption{\label{materialParams}
Material parameters used in the calculations, taken from Ref. \onlinecite{Madelung1982} except where noted.
The gaps are the unstrained indirect gaps appropriate for each material, $E^\Delta_g$ for Si and $E_g^L$ for Ge.
$E_v$ is the unstrained valence energy. }
\end{table}

\begin{figure}
\includegraphics [width=8.5cm]{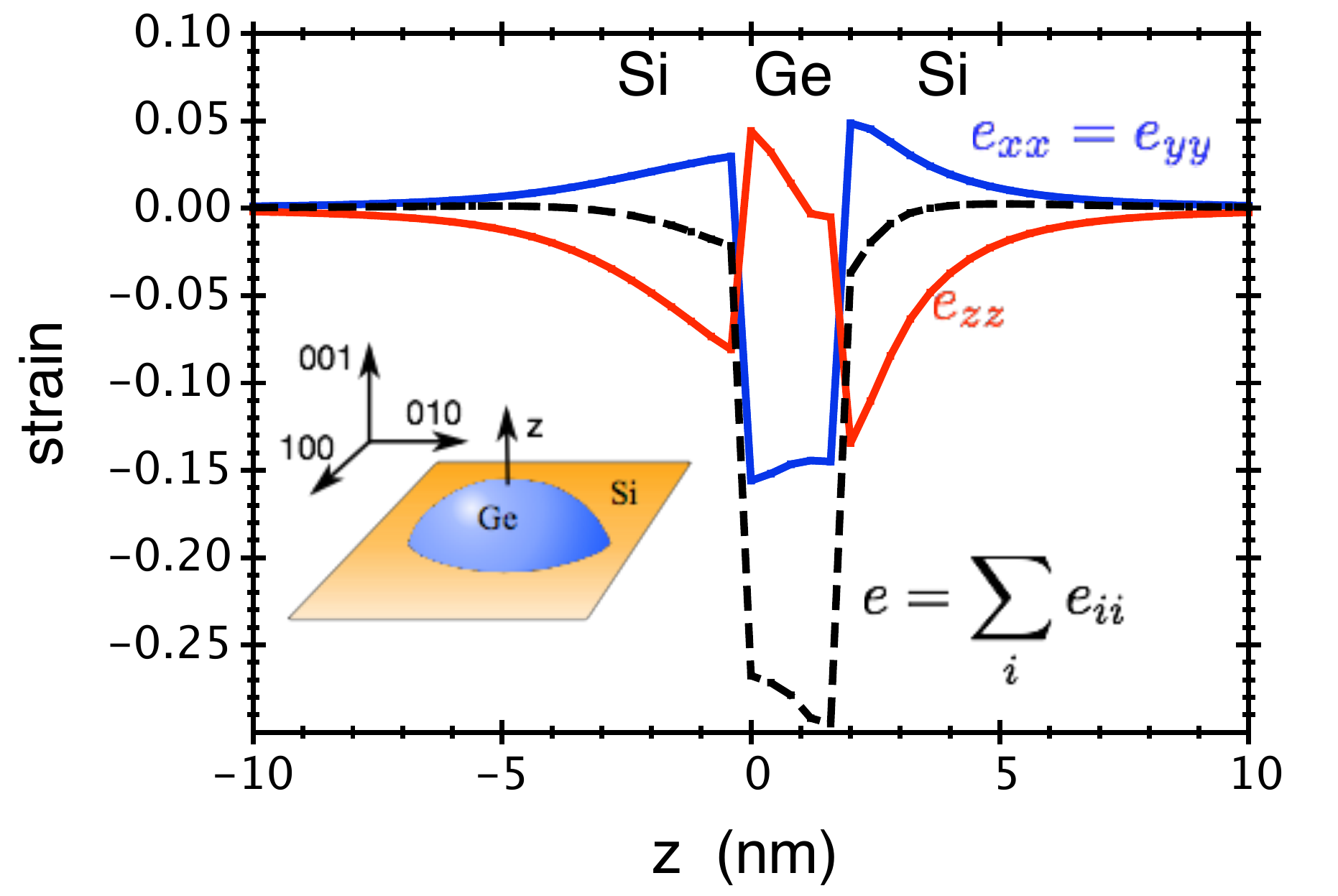}
\caption{\label{fig:1} Strain along a line in the $[001]$ direction through the center of the dot. The strain components are labeled on the curves; the dashed line is the hydrostatic strain.}
\end{figure}

{\it Results}.---Fig.~1 shows the strain for the Ge/Si island we consider, with a height of $2$~nm and a base diameter of $8.4$~nm. The strain components are labeled in Fig. 1, and the dashed line is the hydrostatic strain.
Strain splits $\Delta$-valleys in Si above the Ge dot, with 2 valleys forming wells (here the [001] and [00$\overline{1}$] valleys), and 4 forming barriers.

\begin{figure}
\includegraphics [width=8.5cm]{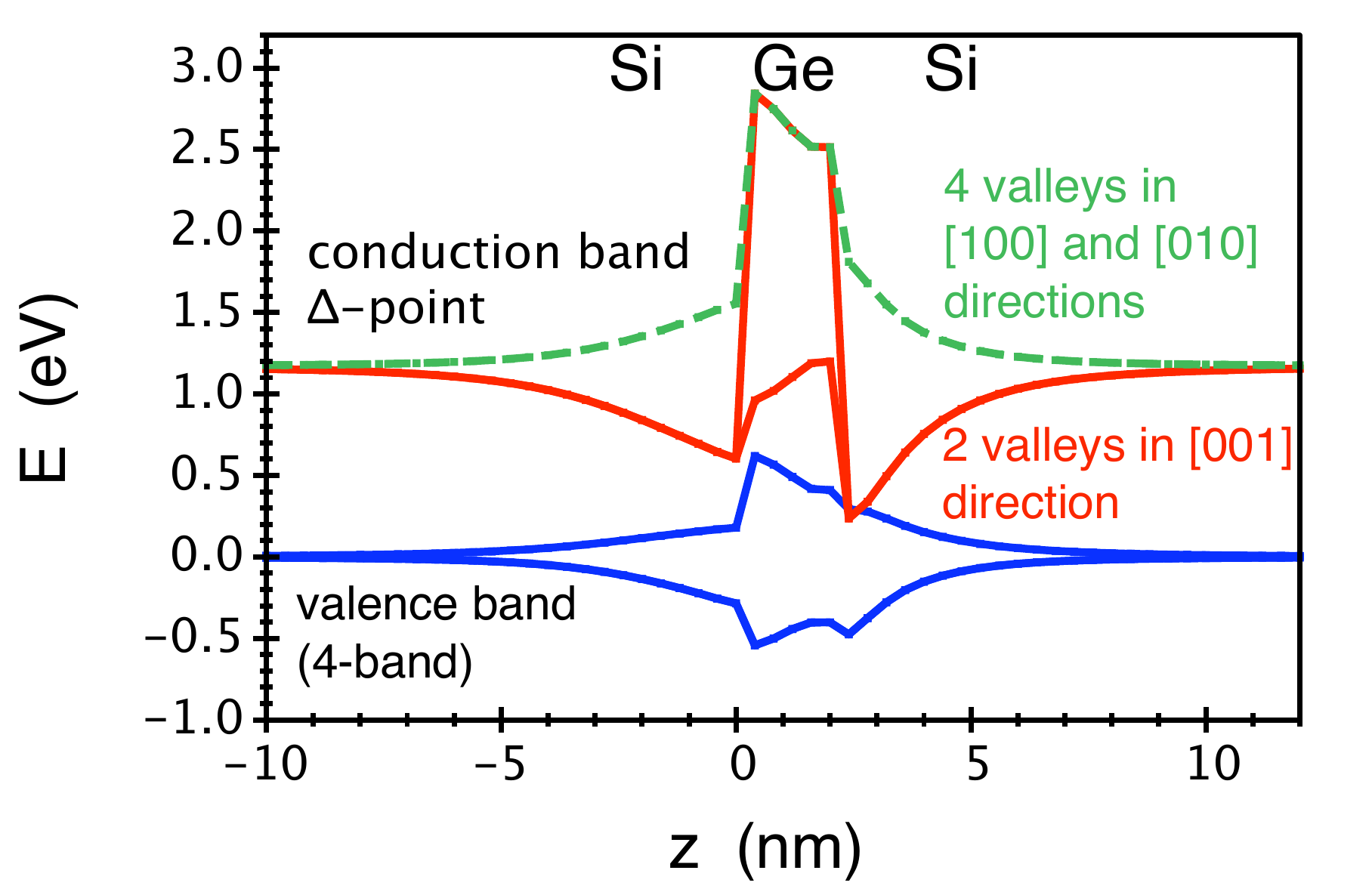}
\caption{\label{fig:2} Band structure of a Ge/Si quantum dot. Curves indicate the k=0 energies for bulk materials having the same strain as a given point in the structure. 
The conduction band states are for the $\Delta$-point. }
\end{figure}

Figure 2 shows the local strain-induced band structure, which is computed at each point in the structure as the band energies that would result in bulk material experiencing the same value of the strain as at that point in the dot structure.  The graph shows those energies along a line in the $[001]$ direction through the center of the Ge dot. 
For the valleys along the $[001]$ direction the  strain induces a confining potential $0.5~ \rm eV$ deep below the dot and $0.9 ~\rm eV$ deep above the dot. Fig. 2 shows that spatially indirect band edges result from the strain, in which holes are confined to the Ge dot and electrons are confined to the Si material surrounding the dot. The wave functions in the absence of electric field are shown in Fig.~3; the holes are confined to the Ge dot, and the electrons are confined to the strain-induced pocket in Si above the dot. 
From the strain shown in Fig. 1 and the conduction band edge shift given in Eq.~(\ref{HDstrain})  we see that the strain above the Ge island shifts the conduction energy downward for the valleys in the [001] and [00$\overline{1}$] direction, while the valleys in the other directions are shifted upward.

\begin{figure}
\includegraphics [width=6.5cm]{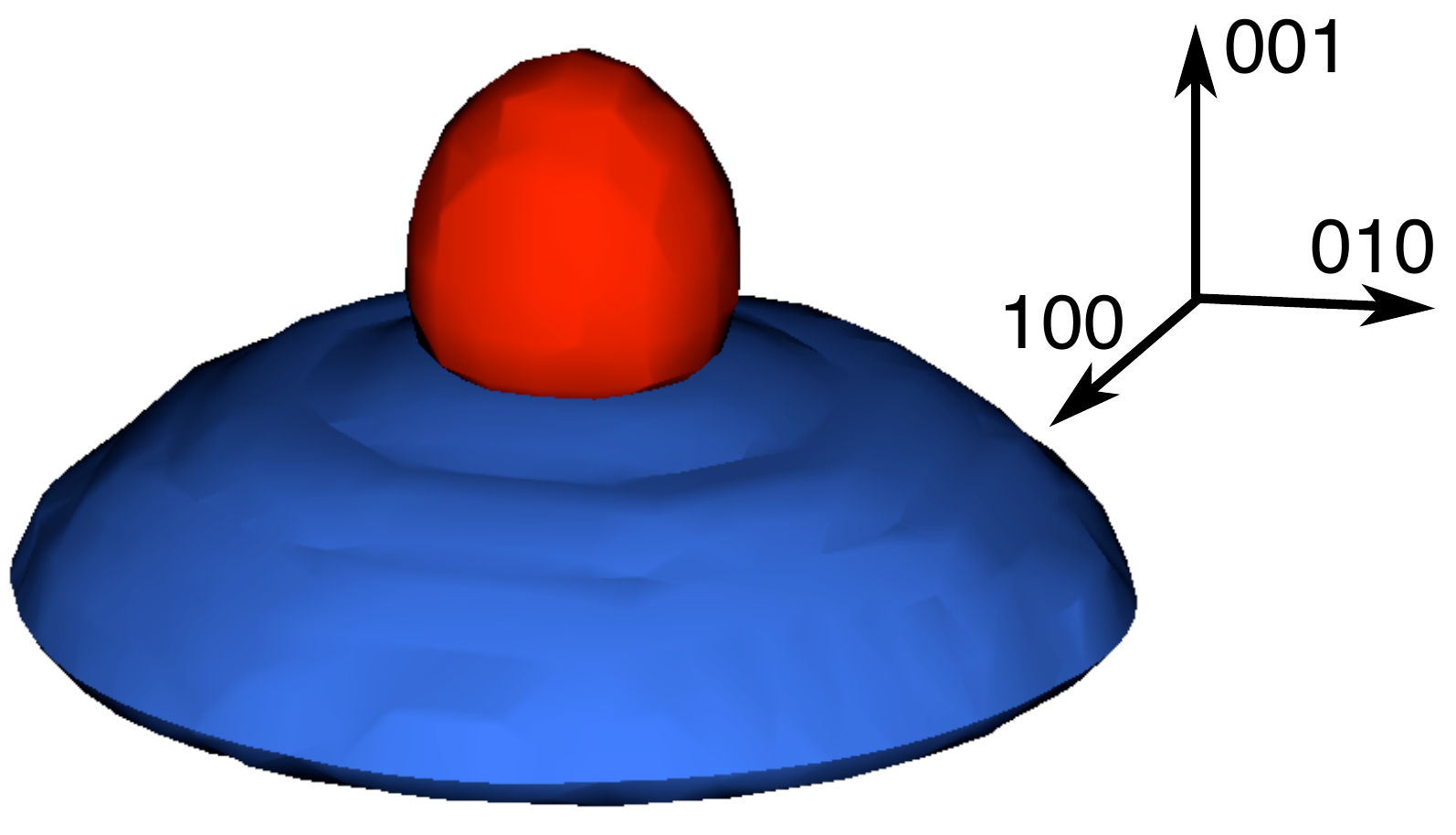}
\caption{\label{fig:3} Contours of $|\psi^2|$ for electron (red/light) and hole (blue/dark) in the absence of an electric field. }
\end{figure}

{\it Field Dependence}.---The existence of distinct potential wells above and below the Ge island suggests the possibility of manipulating the electronic potential with an applied field.
With a sufficiently large electric field the potential well below the dot will have a lower energy.
The insets in Fig. 4 show $|\psi ({\bf  r} ) |^2$ as a function of the electric field.  At small applied fields the electron is localized in the well above the Ge island (ALS).  For larger fields the electron is localized in the well below the Ge island (BLS). The switch of the ground state from ALS to BLS occurs for a change of electric field strength of merely 0.1 kV/cm. If the dots can be confined to a region 10~nm thick, that change in field  corresponds to a 0.1~mV change in voltage.  At finite temperature, however, both the ALS and BLS will be occupied unless the energy difference between the two exceeds $k_BT$.  Thus, for temperatures above $\sim 1$K a larger field will be required to effectively switch from ALS to BLS. For the dot in Fig.~2, at room temperature, which has a distance of $\sim 3$~nm from the peak of the ALS to the peak of the BLS, a shift of $\sim 100$~kV/cm would be effective. Such a field is still below the breakdown field of this material.

\begin{figure}
\includegraphics [width=8.5cm]{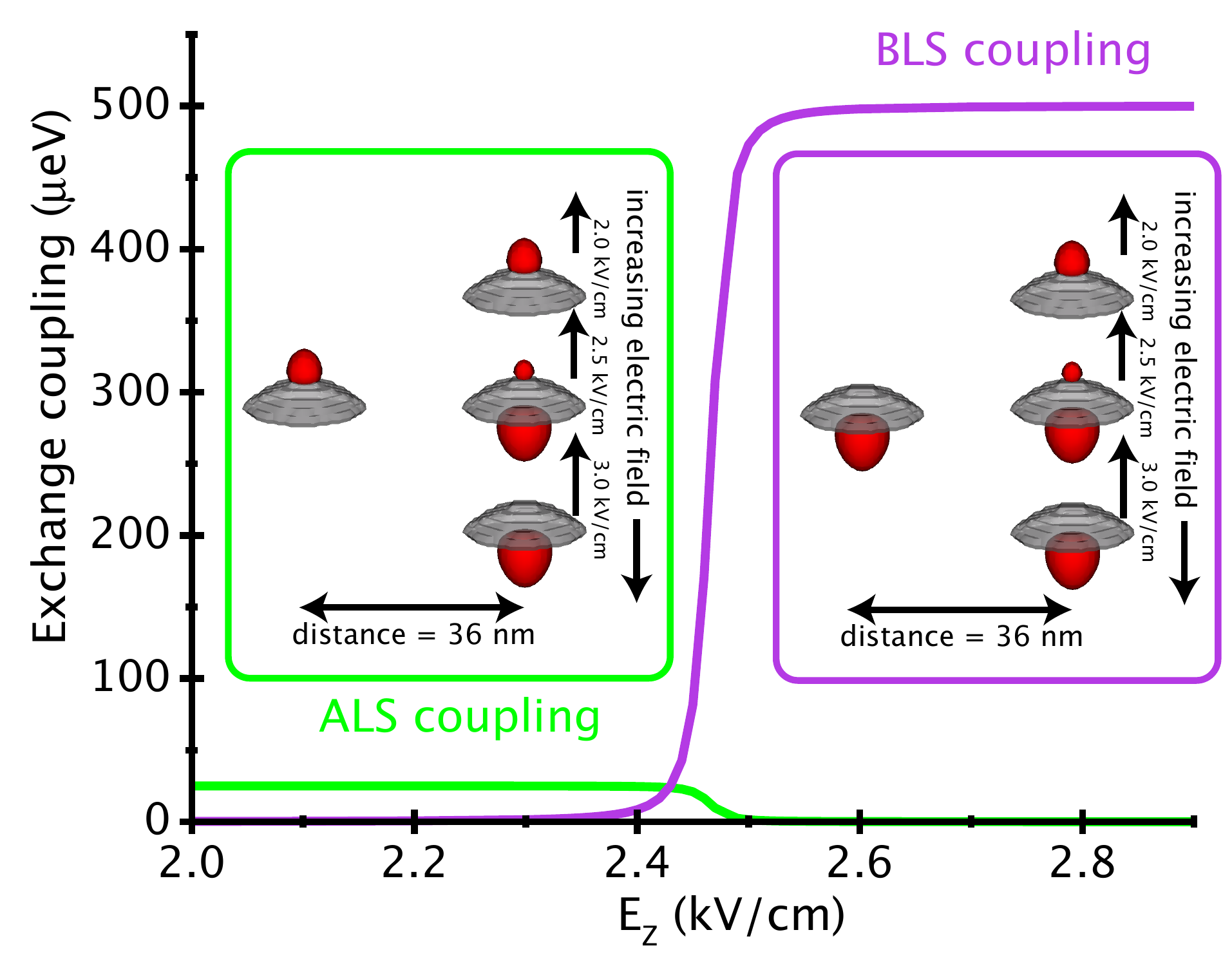}
\caption{\label{fig:4} Electric-field dependence of the exchange coupling of one dot to another dot with an ALS located 36 nm away (green, light) and to another dot with a BLS located 36 nm away (purple, dark). Inset images show the  change in electron localization from above the dot at $2.0~\rm kV/cm$ to below the dot at $3.0~\rm kV/cm$. }
\end{figure}

As a small field shifts the electron from an ALS to a BLS, all electronic properties of the dot can be expected to change rapidly with an external electric field. As a specific example we estimate the exchange interaction between two electron spins confined to two different dots, either in ALS or BLS. When the electrons are both in ALS the lateral decay length is 6~nm, whereas when the electrons are both in BLS the lateral decay length is 12~nm. The coefficient of the exchange decay is 10~meV for both ALS and BLS. Thus for dots 36~nm apart the exchange coupling is $500$~$\mu$eV for BLS and $25$~$\mu$eV for ALS. There is a negligible overlap of an ALS and a BLS. Thus, as shown in Fig.~4, if one dot has an electron in an ALS the exchange coupling can be changed from $25$~$\mu$eV to $0$ with a small electric field change, whereas if one dot has an electron in a BLS the exchange coupling can be changed from $500$~$\mu$eV to $0$ with a small electric field. This provides a quasi-digital approach to controlling the exchange interaction between two electrons confined to two quantum dots.

{\it Conclusions}  We find that the lowest-energy conduction electron states in Si/Ge quantum dots are either apex localized or base localized, depending on the value of a small applied electric field. The localization potential in the Si comes primarily from the inhomogeneous strain induced by the dot. A small electric field can shift the electrons from an ALS to a BLS reversibly, thus modifying the material properties of the dots.   For laterally coupled dots, individually gated dots could have their exchange interactions controlled in a quasi-digital fashion, modified from $500$~$\mu$eV to $0$ with a change of voltage of only $\sim 0.1$~mV.

{\it Acknowledgments} We acknowledge support from DARPA/ARO, an ONR MURI, an NSF NIRT and DOE BES.

\vfill\eject


\begin{thebibliography}{11}
\expandafter\ifx\csname natexlab\endcsname\relax\def\natexlab#1{#1}\fi
\expandafter\ifx\csname bibnamefont\endcsname\relax
  \def\bibnamefont#1{#1}\fi
\expandafter\ifx\csname bibfnamefont\endcsname\relax
  \def\bibfnamefont#1{#1}\fi
\expandafter\ifx\csname citenamefont\endcsname\relax
  \def\citenamefont#1{#1}\fi
\expandafter\ifx\csname url\endcsname\relax
  \def\url#1{\texttt{#1}}\fi
\expandafter\ifx\csname urlprefix\endcsname\relax\def\urlprefix{URL }\fi
\providecommand{\bibinfo}[2]{#2}
\providecommand{\eprint}[2][]{\url{#2}}

\bibitem[{\citenamefont{Bimberg et~al.}(1998)\citenamefont{Bimberg, Grundmann,
  and Ledentsov}}]{Bimberg1998}
\bibinfo{author}{\bibfnamefont{D.}~\bibnamefont{Bimberg}},
  \bibinfo{author}{\bibfnamefont{M.}~\bibnamefont{Grundmann}},
  \bibnamefont{and} \bibinfo{author}{\bibfnamefont{N.~N.}
  \bibnamefont{Ledentsov}}, \emph{\bibinfo{title}{Quantum Dot
  Heterostructures}} (\bibinfo{publisher}{Wiley}, \bibinfo{address}{New York},
  \bibinfo{year}{1998}).

\bibitem[{\citenamefont{Hanson et~al.}(2007)\citenamefont{Hanson, Kouwenhoven,
  Petta, Tarucha, and Vandersypen}}]{Hanson2007}
\bibinfo{author}{\bibfnamefont{R.}~\bibnamefont{Hanson}},
  \bibinfo{author}{\bibfnamefont{L.~P.} \bibnamefont{Kouwenhoven}},
  \bibinfo{author}{\bibfnamefont{J.~R.} \bibnamefont{Petta}},
  \bibinfo{author}{\bibfnamefont{S.}~\bibnamefont{Tarucha}}, \bibnamefont{and}
  \bibinfo{author}{\bibfnamefont{L.~M.~K.} \bibnamefont{Vandersypen}},
  \bibinfo{journal}{Reviews of Modern Physics} \textbf{\bibinfo{volume}{79}},
  \bibinfo{eid}{1217} (\bibinfo{year}{2007}).

\bibitem[{\citenamefont{Ciorga et~al.}(2000)\citenamefont{Ciorga, Sachrajda,
  Hawrylak, Gould, Zawadzki, Jullian, Feng, and Wasilewski}}]{Ciorga2000}
\bibinfo{author}{\bibfnamefont{M.}~\bibnamefont{Ciorga}},
  \bibinfo{author}{\bibfnamefont{A.~S.} \bibnamefont{Sachrajda}},
  \bibinfo{author}{\bibfnamefont{P.}~\bibnamefont{Hawrylak}},
  \bibinfo{author}{\bibfnamefont{C.}~\bibnamefont{Gould}},
  \bibinfo{author}{\bibfnamefont{P.}~\bibnamefont{Zawadzki}},
  \bibinfo{author}{\bibfnamefont{S.}~\bibnamefont{Jullian}},
  \bibinfo{author}{\bibfnamefont{Y.}~\bibnamefont{Feng}}, \bibnamefont{and}
  \bibinfo{author}{\bibfnamefont{Z.}~\bibnamefont{Wasilewski}},
  \bibinfo{journal}{Phys. Rev. B} \textbf{\bibinfo{volume}{61}},
  \bibinfo{pages}{R16315} (\bibinfo{year}{2000}).

\bibitem[{\citenamefont{Simmons et~al.}(2007)\citenamefont{Simmons, Thalakulam,
  Shaji, Klein, Qin, Blick, Savage, Lagally, Coppersmith, and
  Eriksson}}]{Simmons2007}
\bibinfo{author}{\bibfnamefont{C.~B.} \bibnamefont{Simmons}},
  \bibinfo{author}{\bibfnamefont{M.}~\bibnamefont{Thalakulam}},
  \bibinfo{author}{\bibfnamefont{N.}~\bibnamefont{Shaji}},
  \bibinfo{author}{\bibfnamefont{L.~J.} \bibnamefont{Klein}},
  \bibinfo{author}{\bibfnamefont{H.}~\bibnamefont{Qin}},
  \bibinfo{author}{\bibfnamefont{R.~H.} \bibnamefont{Blick}},
  \bibinfo{author}{\bibfnamefont{D.~E.} \bibnamefont{Savage}},
  \bibinfo{author}{\bibfnamefont{M.~G.} \bibnamefont{Lagally}},
  \bibinfo{author}{\bibfnamefont{S.~N.} \bibnamefont{Coppersmith}},
  \bibnamefont{and} \bibinfo{author}{\bibfnamefont{M.~A.}
  \bibnamefont{Eriksson}}, \bibinfo{journal}{Appl. Phys. Lett.}
  \textbf{\bibinfo{volume}{91}}, \bibinfo{pages}{213103}
  (\bibinfo{year}{2007}).

\bibitem[{\citenamefont{Schmidt et~al.}(2000)\citenamefont{Schmidt, Eberl, and
  Rau}}]{Schmidt.prb.2000}
\bibinfo{author}{\bibfnamefont{O.~G.} \bibnamefont{Schmidt}},
  \bibinfo{author}{\bibfnamefont{K.}~\bibnamefont{Eberl}}, \bibnamefont{and}
  \bibinfo{author}{\bibfnamefont{Y.}~\bibnamefont{Rau}},
  \bibinfo{journal}{Phys. Rev. B} \textbf{\bibinfo{volume}{62}},
  \bibinfo{pages}{16715} (\bibinfo{year}{2000}).

\bibitem[{\citenamefont{Loss and DiVincenzo}(1998)}]{Loss1998}
\bibinfo{author}{\bibfnamefont{D.}~\bibnamefont{Loss}} \bibnamefont{and}
  \bibinfo{author}{\bibfnamefont{D.~P.} \bibnamefont{DiVincenzo}},
  \bibinfo{journal}{\pra} \textbf{\bibinfo{volume}{57}}, \bibinfo{pages}{120}
  (\bibinfo{year}{1998}).

\bibitem[{\citenamefont{Kane}(1998)}]{Kane1998}
\bibinfo{author}{\bibfnamefont{B.~E.} \bibnamefont{Kane}},
  \bibinfo{journal}{Nature} \textbf{\bibinfo{volume}{393}},
  \bibinfo{pages}{133} (\bibinfo{year}{1998}).

\bibitem[{\citenamefont{Levy}(2001)}]{Levy2001}
\bibinfo{author}{\bibfnamefont{J.}~\bibnamefont{Levy}}, \bibinfo{journal}{Phys.
  Rev. A} \textbf{\bibinfo{volume}{64}}, \bibinfo{pages}{052306}
  (\bibinfo{year}{2001}).

\bibitem[{\citenamefont{Pryor et~al.}(1997)\citenamefont{Pryor, Pistol, and
  Samuelson}}]{Pryor1997}
\bibinfo{author}{\bibfnamefont{C.}~\bibnamefont{Pryor}},
  \bibinfo{author}{\bibfnamefont{M.-E.} \bibnamefont{Pistol}},
  \bibnamefont{and}
  \bibinfo{author}{\bibfnamefont{L.}~\bibnamefont{Samuelson}},
  \bibinfo{journal}{Phys. Rev. B} \textbf{\bibinfo{volume}{56}},
  \bibinfo{pages}{10404} (\bibinfo{year}{1997}).

\bibitem[{\citenamefont{Fischetti and Laux}(1996)}]{Fischetti.jap.1996}
\bibinfo{author}{\bibfnamefont{M.}~\bibnamefont{Fischetti}} \bibnamefont{and}
  \bibinfo{author}{\bibfnamefont{S.}~\bibnamefont{Laux}}, \bibinfo{journal}{J.
  Appl. Phys.} \textbf{\bibinfo{volume}{80}}, \bibinfo{pages}{2234}
  (\bibinfo{year}{1996}).

\bibitem[{\citenamefont{Madelung et~al.}(1982)\citenamefont{Madelung, Schilz,
  and Weiss}}]{Madelung1982}
\bibinfo{editor}{\bibfnamefont{O.}~\bibnamefont{Madelung}},
  \bibinfo{editor}{\bibfnamefont{M.}~\bibnamefont{Schilz}}, \bibnamefont{and}
  \bibinfo{editor}{\bibfnamefont{H.}~\bibnamefont{Weiss}}, eds.,
  \emph{\bibinfo{title}{Landolt-B{\"o}rnstein, Numerical Data and Functional
  Relations in Science and Technology}} (\bibinfo{publisher}{Springer},
  \bibinfo{address}{New York}, \bibinfo{year}{1982}).

\end{thebibliography}
\end{document}